\documentstyle[aps,epsfig,preprint]{revtex}
\begin{document}
\vspace{1.5cm}
\begin{center}
{\Large \bf
$\Omega$, $J/\psi$ and $\psi^{\prime}$ Transverse Mass Spectra at RHIC 
} 

\vspace{0.5cm}
{\bf K.A. Bugaev}$^{a,b}$,
{\bf M. Ga\'zdzicki}$^c$ and
{\bf M.I. Gorenstein}$^{a,d}$ 

\end{center}

\vspace{0.3cm}
\noindent
$^a$ Bogolyubov Institute
for Theoretical Physics,
Kiev, Ukraine\\
$^b$ Gesellschaft f\"ur Schwerionenforschung (GSI), Darmstadt, Germany\\
$^c$ Institut f\"ur  Kernphysik, Universit\"at  Frankfurt,
Germany\\
$^d$ Institut f\"ur Theoretische Physik, Universit\"at  Frankfurt,
Germany\\

\vspace*{-1.0cm}

\begin{abstract}
\noindent
The transverse mass spectra of $J/\psi$ and $\psi'$ mesons
and $\Omega$ hyperons 
produced in central Au+Au collisions at RHIC energies
are discussed within a statistical 
model used successfully for the interpretation of the SPS results.
The comparison of the presented model
with the future RHIC data 
should serve as a further crucial
test of the hypothesis of statistical production of charmonia
at hadronization. 
Finally, in  case of validity, the approach should 
allow to estimate the mean transverse flow velocity at 
the quark gluon plasma hadronization. 
\end{abstract}

\vspace{0.5cm}
The  concepts of chemical (hadron multiplicities) and kinetic
(hadron momentum spectra) freeze--outs were introduced to interpret data on
hadron production in relativistic nucleus--nucleus (A+A) collisions.
The equilibrium hadron gas (HG) model describes remarkably well 
the light hadron
multiplicities measured in A+A collisions at the SPS \cite{HG}
and RHIC \cite{HG1} energies, where the creation of a quark--gluon plasma
(QGP) is expected. 
Recently it was found \cite{gago} that also  charmonium 
yield systematics in nuclear collisions at
the SPS \cite{na50} follow the pattern predicted by the hadron gas model.
The
hadronization temperature parameter extracted from the fits to the
hadron multiplicities is
similar for both energies: $T_H=170\pm 10$~MeV.
It is close to an
estimate of the temperature $T_C$ for
the
QGP--HG transition at zero baryonic density.

Experimental results on inclusive hadron spectra and
correlations  
show evidence for a hydrodynamic expansion of the matter
created in heavy ion collisions.
Strong transverse flow effects in Pb+Pb collisions at the SPS (average
collective transverse velocity is approximately $0.5\div 0.6$) are firmly
established from the combined analysis
\cite{ka,Wi:99} of the pion transverse mass spectra and
correlations.
The kinetic (`thermal') freeze--out of pions and nucleons seems to
occur at a rather late
stage of an A+A reaction. 
The thermal freeze--out temperature parameter of pions
measured by the NA49 Collaboration \cite{na49} for central Pb+Pb collisions 
at the SPS is 
$T_f\cong 120$~MeV.

Further exploration of  idea of the statistical 
$J/\psi$ production \cite{gago} led to the formulation of the
hypothesis that the kinetic freeze--out of $J/\psi$  and $\psi^{\prime}$ 
mesons takes place directly at hadronization \cite{BGG}.
This means that  chemical and thermal
freeze--outs occur simultaneously for those mesons and they, therefore,
carry information on the flow velocity of strongly interacting matter just
after
the  transition to the HG phase.
A possible influence of
the effect of rescattering 
in the hadronic phase on the transverse momentum ($p_T$) spectra
was recently studied
within 
a ``hydro + cascade'' approach \cite{BD,Sh}.
A+A collisions are considered there to split into three stages:
hydrodynamic QGP
expansion (``hydro''),
transition from QGP to HG (``switching'')
and the stage of hadronic rescattering and resonance decays
(``cascade'').
The switching from hydro to cascade takes place at $T=T_C$, where the
spectrum of hadrons leaving the surface of the QGP--HG transition is taken
as an input for the subsequent cascade calculations.
The results \cite{BD,Sh} suggest that
the $p_T$ spectrum of $\Omega$s is only weakly affected during
the cascade stage. The corresponding calculation for charmonia
are not yet performed within this model, but a similar result may be 
expected due to their very high masses and low interaction cross sections. 

In previous work \cite{Go:02} devoted to the analysis of the SPS data
 it was demonstrated that the  measured  
transverse mass ($m_T=\sqrt{p_T^2 + m^2}$)  spectra of
$\Omega^{\pm}$ hyperons \cite{Omega} and 
$J/\psi$  and $\psi^{\prime}$ mesons \cite{NA50}
produced in Pb+Pb  at 
158~A$\cdot$GeV  collisions
 can be reproduced within a hydrodynamical approach
using the same 
freeze--out parameters: hadronization temperature
$T\cong 170$~MeV and the mean transverse flow
velocity  $\overline{v}_T\cong
0.2$.
Within our approach the value of the $\overline{v}_T$
parameter extracted in this way should be interpreted as
the mean flow velocity of the hadronizing QGP.

\vspace{0.3cm} 
In the present letter we discuss
the transverse mass spectra of 
$J/\psi$  and $\psi^{\prime}$ mesons  and $\Omega$ hyperons in central Au+Au
collisions at RHIC energies within the statistical approach to
charmonium production successfully used for the interpretation
of the SPS data \cite{gago,Go:02}.

Assuming kinetic
freeze--out of matter
 at constant temperature $T$,
the transverse mass spectrum 
in cylindrically symmetric and longitudinally boost invariant fluid
expansion can be approximated as \cite{Heinz}:
\begin{equation}\label{hydro}
\frac{dN_i}{m_T dm_T}~
\propto~
m_T~ \int_{0}^{R}r dr~ K_1\left({ \frac{m_T \cosh y_T}{T} }\right)~  
I_0\left({ \frac{p_T\sinh y_T}{T}}\right)~,
\end{equation}
where $y_T=\tanh^{-1}v_T$ is the transverse fluid rapidity, $R$ is the
transverse system size, $K_1$ and $I_0$ are the modified Bessel functions.
The spectrum (\ref{hydro}) is obtained under the assumption that the
freeze--out occurs at constant longitudinal proper time $\tau 
=\sqrt{t^2-z^2}$ 
($t$ is the time and $z$ is the longitudinal coordinate), i.e. the
freeze--out time $t$ is independent of the transverse coordinate $r$.  
The analysis of the numerical 
calculations done within a hydrodynamical model \cite{Sh}
shows that the latter is approximately fulfilled.
The quality of the approximation made gets better for   
heavy particles considered here
because a possible
deviation from Eq.(\ref{hydro})
decreases with increasing particle mass at
constant $p_T$.
In order to calculate (\ref{hydro}) the function
$y_T(r)$ has to be given.
A linear flow profile, $y_T(r)=y_T^{max}\cdot r/R$, is often
assumed in phenomenological fits \cite{Heinz} and we used it 
as well.
Recent numerical  hydrodynamical calculations \cite{Sh}
show that deviations from this simplified profile
are smaller than 10\% for both SPS and RHIC energies.
Their importance for our results is discussed later in the
paper.

In Figs.~1 -- 3  the measured 
$m_T$--spectra of 
$\Omega^-$ \cite{Omega},
$J/\psi$ and
$\psi^{\prime}$ \cite{NA50} 
produced in Pb+Pb collisions at 158 A$\cdot$GeV
are shown together with the
fit performed using  Eq.(\ref{hydro}).
The model parameters
are: $T=170$~MeV, $y_T^{max}=0.28$ \cite{Go:02}. 
The value of $y_T^{max}$ found from fitting the SPS data appears to be
close to the numerical estimate done within a hydrodynamical approach
($y_T^{max} \approx 0.3$)
\cite{Sh}. 
This encourages us to use a corresponding estimate of 
$y_T^{max}\cong 0.55$ made for Au+Au collisions at
$\sqrt{s}_{NN}=130$~GeV \cite{Sh} for
the first guess of the $m_T$--spectra of charmonia at RHIC
energies within our approach.
As follows from the analysis 
\cite{HG1} of hadron yields within the HG model,
in our calculations we use a hadronization temperature
$T=170$ MeV,
the same as in the case of the SPS data.
The resulting $m_T$--spectra 
for $\Omega$,  $J/\psi$ and $\psi'$ are shown in Figs.~1, 2 and 3, respectively.
The normalization of the SPS and RHIC spectra is arbitrary.
Due to large transverse flow velocity
of hadronizing QGP at RHIC energies, the particle
spectra are expected to significantly deviate from the
simple exponential form of the $m_T$--distribution:
\begin{equation}\label{T*}
\frac{dN}{m_T dm_T}~
= ~
C~m_T^{1/2}~
\exp\left( -~\frac{m_T}{T^*} \right)~,
\end{equation}
where  $C$ and $T^*$ are normalization and inverse slope
parameters, respectively.  

In order to quantify the deviations of the RHIC spectra from Eq. (2)
we plotted in 
 Fig.~4 the local inverse slope defined as:

\vspace*{-0.4cm}

\begin{equation}\label{T*1}
T^*~\equiv~-~\left[ \frac{d}{dm_T} \ln \left(m_T^{-1/2}\frac{dN}{dm_T^2} 
\right) \right]^{-1}~,
\end{equation}
versus $m_T-m$
for $J/\psi$, $\psi^{\prime}$ and $\Omega$.
The $T^*$ parameter was calculated using Eq.(\ref{hydro})
with the values of parameters used previously for the SPS and RHIC energies.
No significant dependence of $T^*$ is observed in the SPS case
$y^{max}_T = 0.28$, whereas a very strong decrease 
of $T^*$ with increasing $m_T$ in the region $m_T-m < 1$~GeV
is seen for $J/\psi$ and $\psi^{\prime}$ at RHIC
($y^{max}_T = 0.55 $) and the moderate one is expected for $\Omega$ hyperons.
Thus the shape of the $m_T$--spectrum is strongly dependent on the
magnitude of the transverse flow and the mass of the particle.
These characteristic ``hydrodynamical'' features of the spectrum
should allow for a clear test of our approach in the near future.

To be more specific, we follow 
the suggestion of Ref. \cite{Sh}
and
consider two domains $0 < m_T - m < 0.6$ GeV
and $0.6 < m_T - m < 1.6$ GeV of $m_T$--spectra.
Within these $m_T$ domains the $\Omega$,
$J/\psi$ and $\psi^{\prime}$ spectra at RHIC calculated 
according to Eq. (1)
were approximated by Eq. (2) and the  values of $C$ and $T^*$
were found by the maximum likelihood method.
The inverse slopes in low $m_T$ domain 
(long dashed lines in Figs. 1 -- 3)
are larger than the high $m_T$ ones (dashed--dotted lines in Figs. 1 -- 3).
However, the difference between low and high $m_T$ inverse
slopes strongly depends on the particle mass --
for $\Omega$ the inverse slopes differ by 13 per cent only, whereas for
$\psi^{\prime}$ by about 40 per cent.

It is important to check the sensitivity of the above results
to the assumed flow profiles (density and velocity) of
hadronic matter at freeze--out.
We first note that because of the basic assumption of our model
($\Omega$, $J/\psi$ and $\psi^{\prime}$ freeze--out at QGP
hadronization) the 
freeze-out density at the boundary 
between the mixed and hadronic phases
is determined by the hadronization temperature
$T = 170$ MeV and hence the density profiles of these particles
are  constant as the function of $r$.  
Thus the uncertainty is due to the velocity profile only.
The numerical hydrodynamical calculations \cite{Sh}
indicate that the transverse rapidity profile may deviate
from the linear dependence assumed above by less than 10\%.
In order to estimate a possible influence of these deviations
we calculated the spectra of $\Omega$, $J/\psi$ and $\psi^{\prime}$
using a profile of the form $y_T(r)=y_T^{max}\cdot ( r/R )^{\alpha} $ 
for $\alpha =$ 0.9 and 1.1.
In the calculations we keep the average 
transverse velocity as for $y_T^{max} = 0.55$ and $\alpha = 1$ case. 
It turns out that the shape of the spectra is almost insensitive
to the assumed value of $\alpha$ parameter 
in the whole range of transverse mass for $\Omega$ and
for $ m_T - m > 0.6$ GeV for $J/\psi$ and $\psi^{\prime}$ mesons. 
In the low $m_T$ domain ($ m_T - m < 0.6$ GeV) the fitted inverse
slope parameters for $J/\psi$ and $\psi^{\prime}$ mesons
vary by 3 per cent and 5 per cent, respectively, under
considered variations of the $\alpha$ parameter.
Therefore we conclude - 
the principal results of our analytical model are weakly
affected by the approximations made.


The transverse flow velocity of  hadronizing matter depends
on the energy and centrality of the collision.
Thus, in general, we consider the parameter $y^{max}_T$
as a free parameter, who's value should be extracted from the
fit of Eq.(\ref{hydro}) to the measured spectrum.
Providing the fit is successful, the fitted $y^{max}_T$ may be
treated as an estimate of the transverse 
flow velocity of  hadronizing matter.

The $m_T$ spectra in Au+Au collisions at RHIC
were also calculated within a hydrodynamical ansatz in Ref.~\cite{bf}.
There is however a principal difference between our approach \cite{Go:02}
and the one used in Ref.~\cite{bf}. 
We consider the early kinetic freeze--out for $\Omega$, $J/\psi$ and
$\psi^{\prime}$ only,  whereas in \cite{bf} the simultaneous 
(chemical and thermal) freeze--out was postulated for all hadrons.
This generates a substantial difference in the model predictions 
for the RHIC freeze--out parameters.

The results on charmonium production in central Au+Au
collisions at RHIC energies
are expected to be available very soon.
They will, obviously, allow to test various approaches \cite{v1,v2,v3,v4,v5}
developed in the course of the analysis of the corresponding
SPS data.
In particular, a statistical approach to charmonium production
in  nuclear collisions yields well defined and characteristic
predictions for charmonia multiplicities 
\cite{gago} and the transverse mass spectra discussed in this
letter.
The analysis of the $m_T$--spectra of  $J/\psi$ and $\psi^{\prime}$
mesons and $\Omega$ hyperons  
within our approach should yield a unique information on the
transverse flow velocity of hadronizing QGP in A+A collisions
at RHIC energies.

\vspace*{0.2cm}

\noindent
{\bf  Acknowledgments.}
We thank P. Braun--Munzinger, W. Broniowski,
W. Florkowski and R. Renfordt for discussions and
comments.
The financial supports from the Humboldt Foundation 
and INTAS grant 00--00366 are acknowledged.


\newpage

\begin{figure}[p]
\vspace*{-1.cm}
\epsfig{file=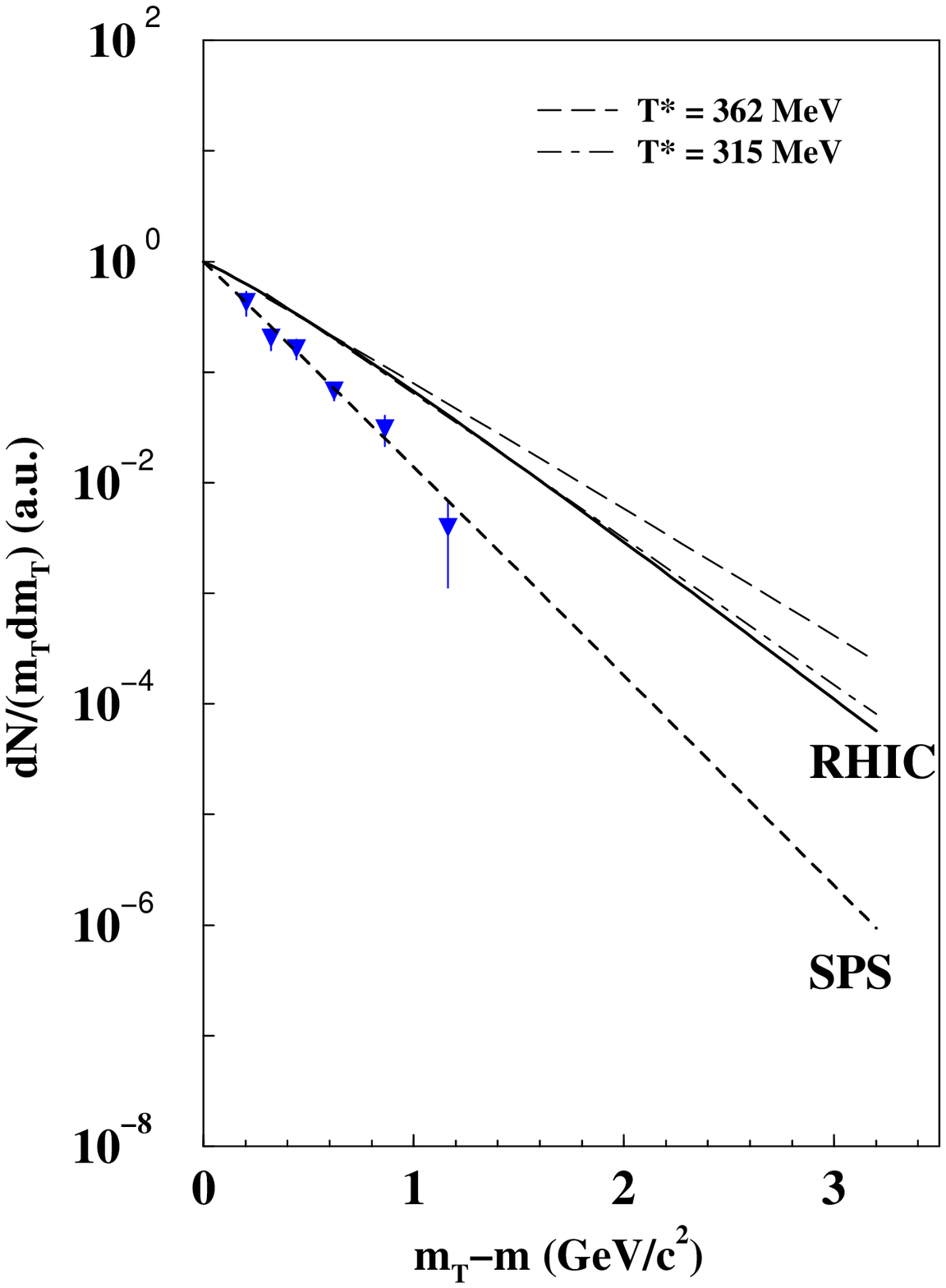,width=14cm}
\caption{
The transverse mass spectra (in arbitrary units) of 
$\Omega^-$ hyperons produced in  Pb+Pb (Au+Au) collisions.
The points indicate experimental results measured for
Pb+Pb collisions at 158 A$\cdot$GeV.
The dashed line shows the fit of the hydrodynamical model
to these data ($T=170$ MeV and $y^{max}_T=0.28$).
The prediction of our approach for central Au+Au
collisions at $\sqrt{s}_{NN}$ = 130 GeV is indicated by the
solid line ($T=170$ MeV and $y^{max}_T=0.55$).
The long dashed and dashed--dotted lines
correspond to the exponential spectrum
fitted in the intervals $(m_T - m) \in [0,\, 0.6]$ GeV and
$(m_T - m) \in [0.6,\, 1.6]$ GeV), respectively.
 } \label{fig1}
\end{figure}

\newpage

\begin{figure}[p]
\epsfig{file=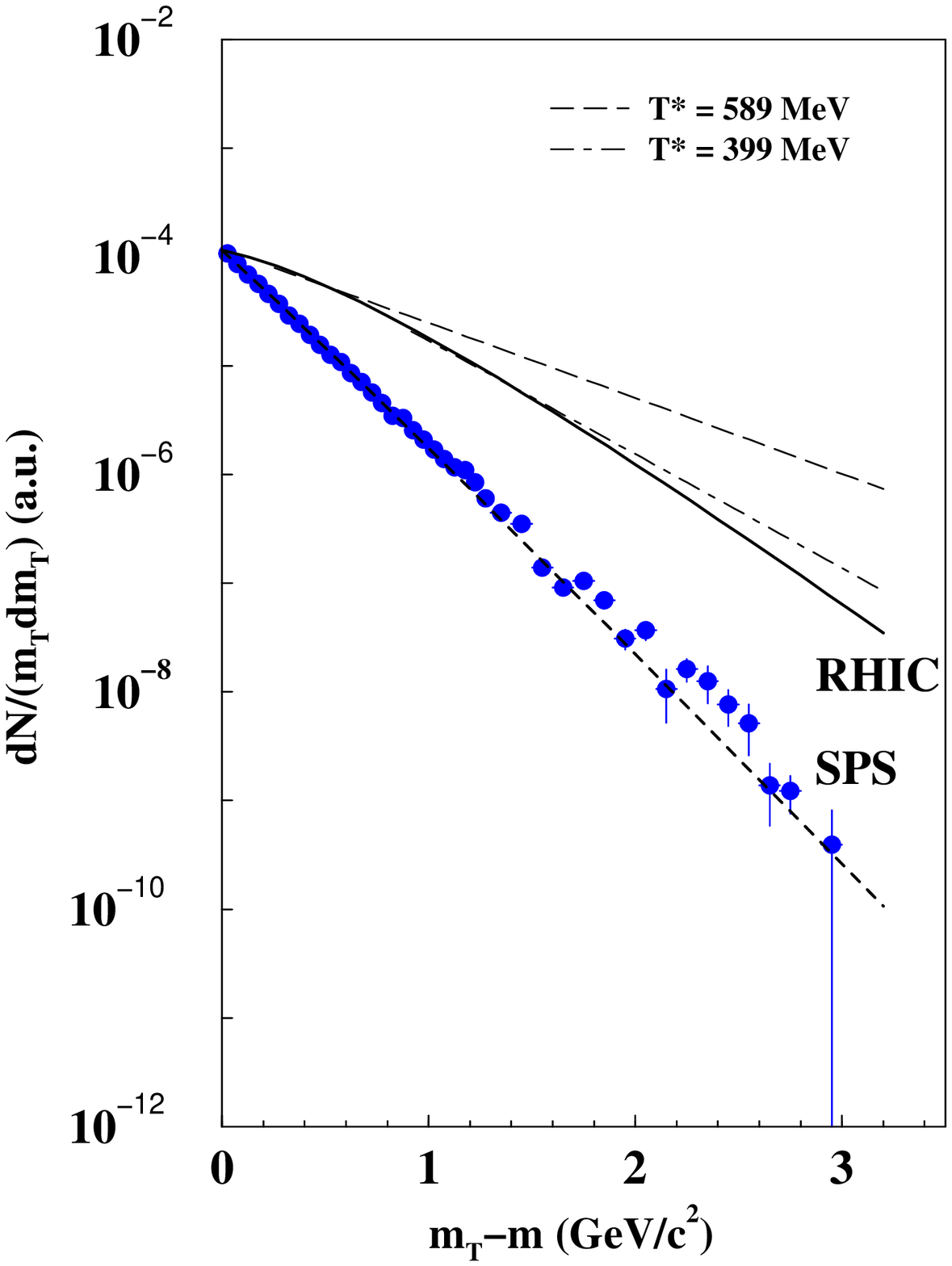,width=14cm}
\caption{
The same as in Fig. \ref{fig1} but for $J/\psi$ mesons.
 } \label{fig2}
\end{figure}

\newpage

\begin{figure}[p]
\epsfig{file=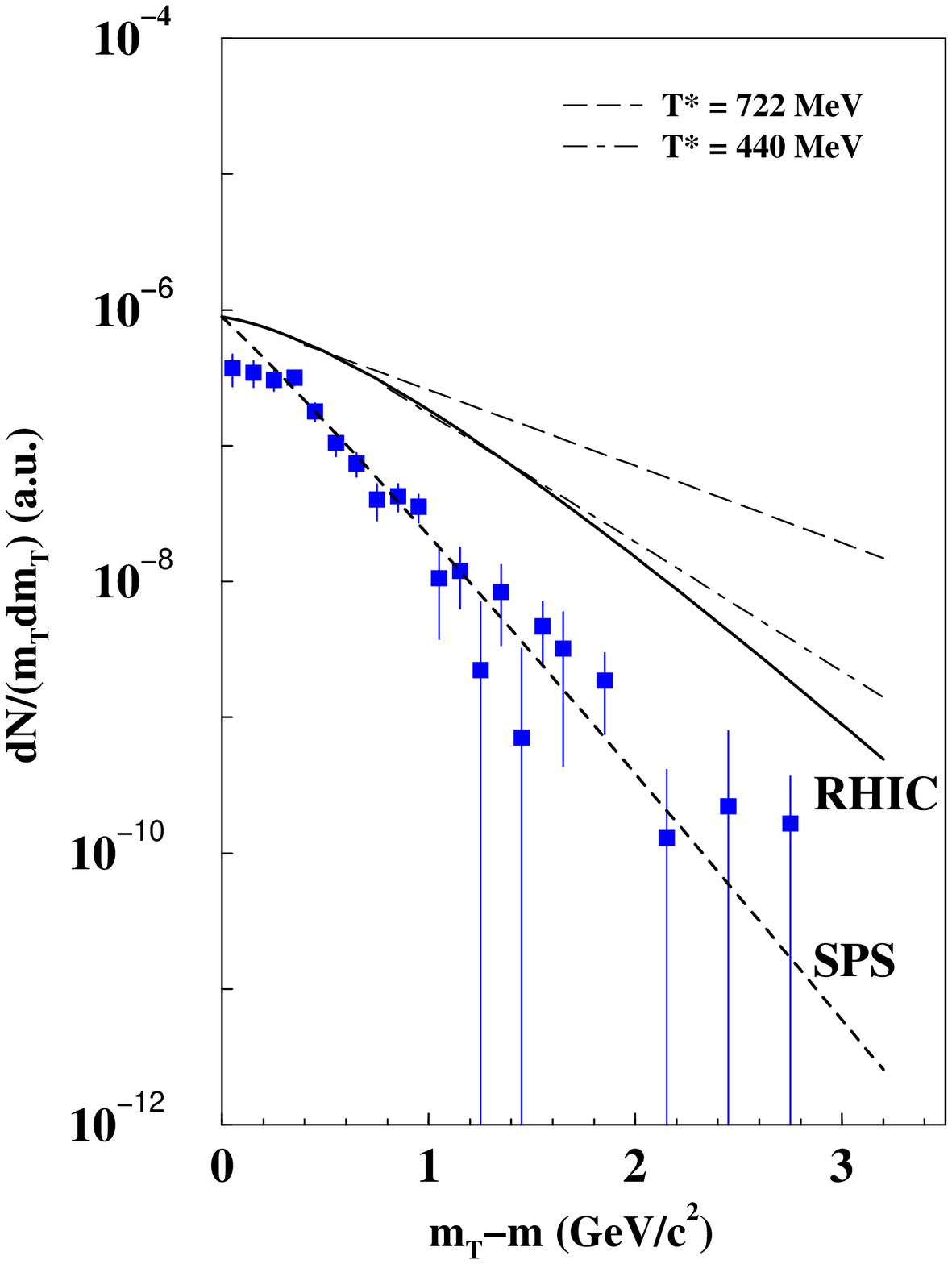,width=14cm}
\caption{
The same as in Fig. \ref{fig1} but for $\psi'$ mesons.
 } \label{fig3}
\end{figure}

\newpage

\begin{figure}[p]
\mbox{\hspace*{-1.0cm}\epsfig{figure=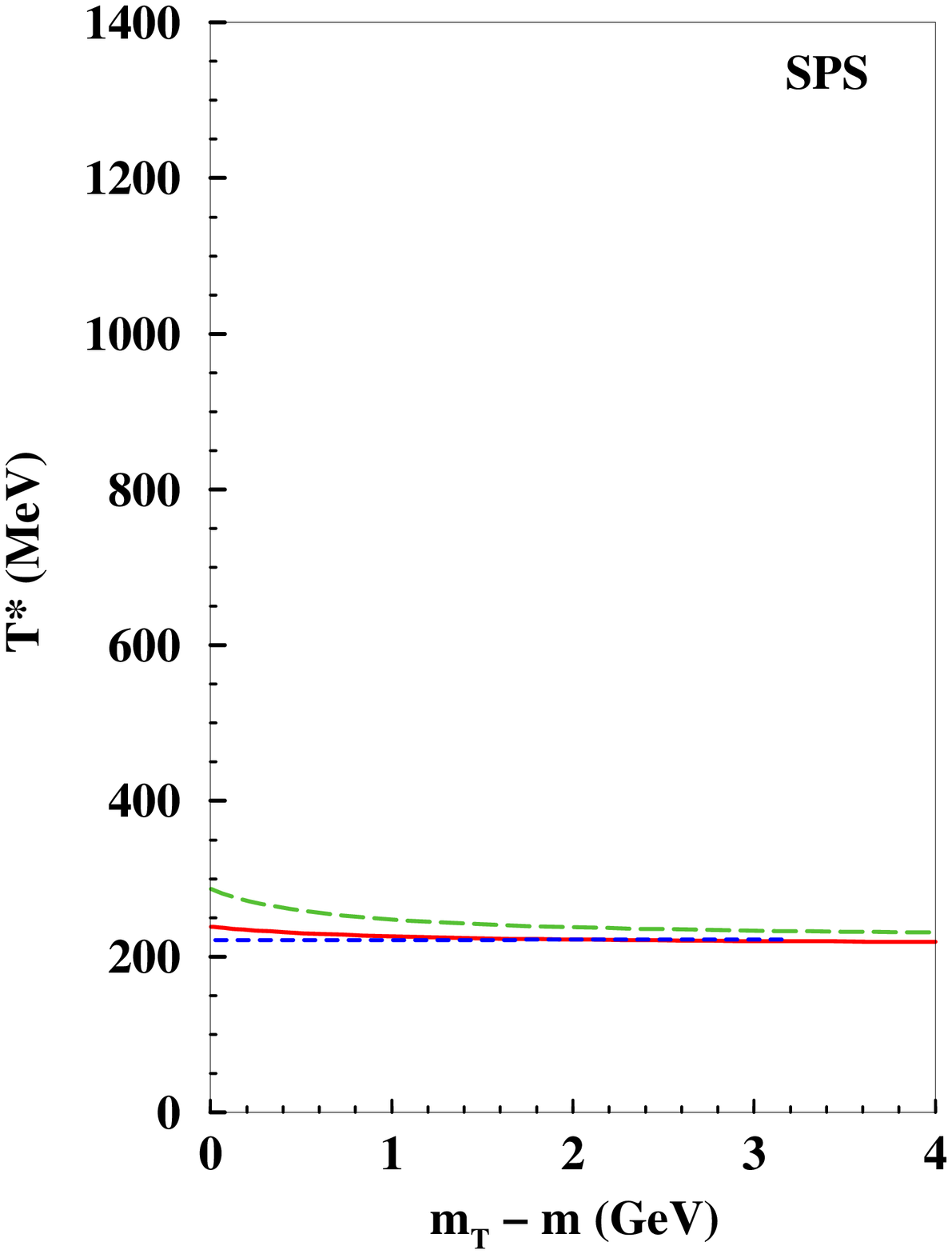,height=12.cm,width=11.5cm}\\
\hspace*{-4.0cm}\epsfig{figure=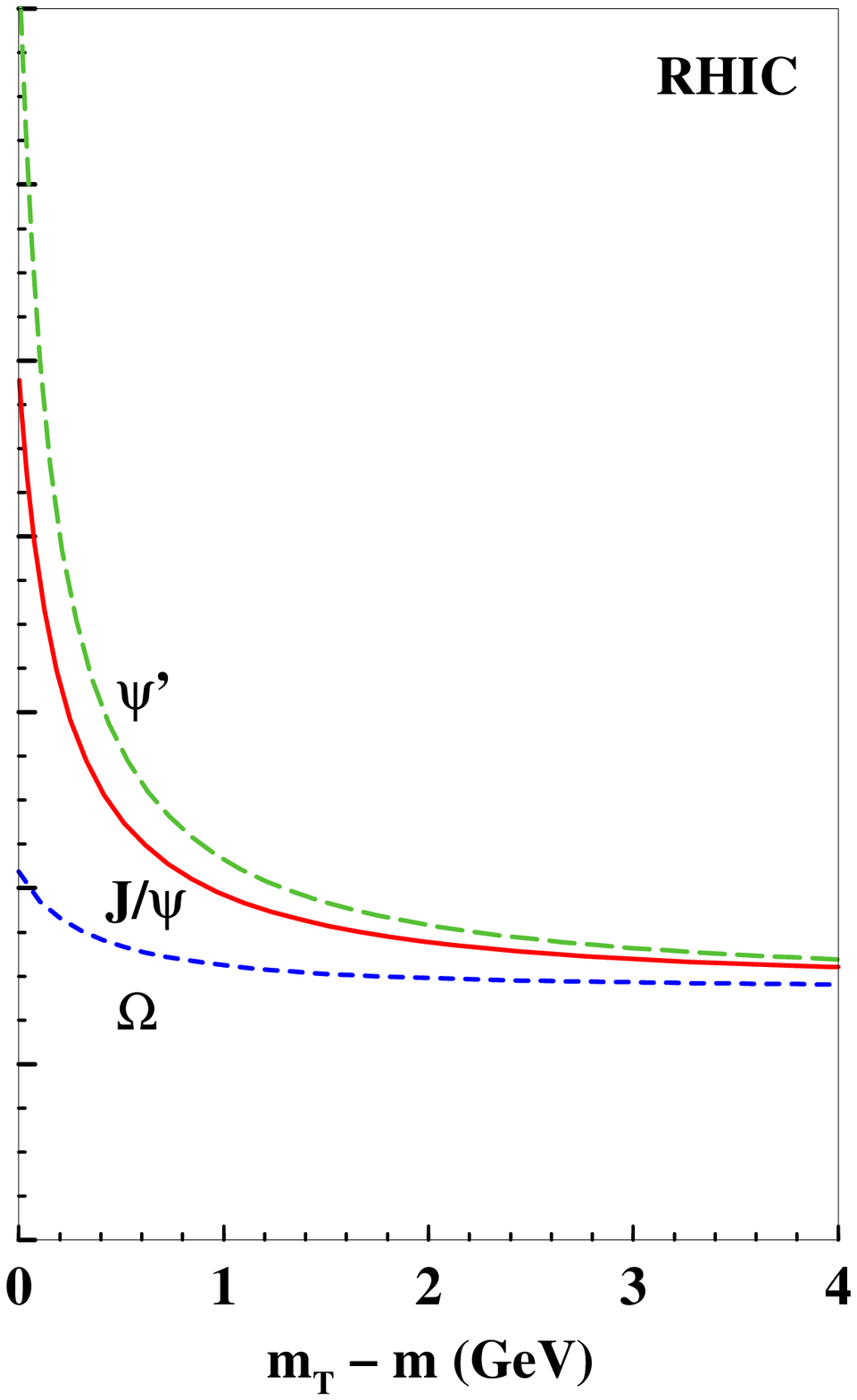,height=12.cm,width=11.5cm}}
\caption{
The dependence of the inverse slope parameter $T^*$ 
of $J/\psi$ (solid line) and $\psi'$ (long dashed line) mesons
and $\Omega$ hyperons (short dashed line) on
$m_T-m$ expected within the model of statistical
production at QGP hadronization.
The left panel indicates the behaviour given by the
fit of the hydrodynamical ansatz to the data on Pb+Pb collisions
at 158 A$\cdot$GeV, whereas the right panel shows
the predictions for central Au+Au collisions at 
$\sqrt{s}_{NN}$ = 130 GeV.
 } \label{fig4}
\end{figure}

\end{document}